\documentclass[aps,pra,showpacs,twocolumn]{revtex4}
\usepackage{amssymb}
\usepackage{graphicx}
\usepackage{amsmath}
\usepackage{subfigure}
\usepackage{epsfig}
\usepackage{multirow}
\usepackage{float}
\usepackage{color}
\usepackage{lscape}
\usepackage{pdflscape}
\usepackage[figuresright]{rotating}
\usepackage{ulem}
\usepackage{lineno}

\begin{document}
\title{Finite field calculations of static polarizabilities and hyperpolarizabilities of In$^{+}$ and Sr}

\author{Yan-mei Yu$^{1}$ \footnote[1]{E-mail: ymyu@aphy.iphy.ac.cn}, Bing-bing Suo$^{2}$\footnote[2]{E-mail: bsuo@nwu.edu.cn}, Hui-hui Feng$^{1}$, Heng Fan$^{1}$, Wu-Ming Liu$^{1}$}
\address{$^1$Beijing National Laboratory for Condensed Matter Physics, Institute of Physics, Chinese Academy of Sciences, Beijing 100190,China}
\address{$^2$Institute of Modern Physics, Northwest University, Xi'an, Shanxi 710069, China}

\date{\today}
\begin{abstract}
The finite field calculations are performed for two heavy frequency-standard candidates In$^+$ and Sr. The progressive hierarchy of electron correlations is implemented by the relativistic coupled-cluster and configuration interaction methods combined with basis set of increasing size. The dipole polarizabilities, dipole hyperpolarizabilities, quadrupole moments, and quadrupole polarizabilities are recommended for the ground state 5s$^2$ $^1S_0$ and low-lying states 5s5p $^3P^{\rm o}_{0,1,2}$ of In$^+$ and Sr. Comparative study of the fully and scalar relativistic electron correlation calculations reveals the effect of the spin-orbit interaction on the dipole polarizabilities of In$^{+}$ and Sr. Finally, the blackbody-radiation shifts due to the dipole polarizability, dipole hyperpolarizability, and quadrupole polarizability are evaluated for the clock transition 5s$^2$ $^1S_0$ - 5s5p $^3P^{\rm o}_0$ of In$^+$ and Sr.

\end{abstract}

\maketitle
\section{Introduction}
Polarizabilities describe the ability of atoms, ions, and molecules to be polarized in an external electric field, which is very useful in many areas of atomic and molecular physics \cite{Mitroy-JPB-2010,Derevianko-RMP-2011,Safronova-IEEE-2012,Schwerdtfeger-web-2012}. In recent years, optical lattice clocks have shown a tremendous progress \cite{Chou-PRL-2010,Huang-PRA-2011,Hinkley-Science-2014,Ye-NC-2015}, which has stimulated a great deal of interest in performing precision calculations of atomic polarizabilities and hyperpolarizabilities. For instance, the difference of the static dipole polarizabilities between two associated states of an atomic clock transition determines the blackbody-radiation (BBR) shift that is crucial in evaluating the error of an atomic clock, which has been calculated for Al$^+$, Ca$^+$, Sr, Yb, In$^+$, Sr$^+$, Hg$^+$, Mg, Ca and so on ~\cite{Mitroy-JPB-2010,Safronova-IEEE-2012,Sahoo-PRA-2009,Safronova-PRL-2011,Safronova-PRA-2013-Yb,Safronova-PRA-2013-Sr}. The quadrupole polarizability is another important quantity that is related to the multipolar BBR shift, which has been evaluated for Sr, Ca$^+$, Sr$^+$, and so on \cite{Porsev-PRA-2006,Arora-PRA-2012}. As a higher-order response to the applied electric field, the hyperpolarizability also contributes to the energy shift of an optical frequency standard, being small but not necessarily negligible, which has been investigated by Ovsiannikov, et al~\cite{Ovsiannikov-PRA-2006, Ovsiannikov-PRA-2013}.

Among the current atomic optical clock candidates, an important category is based on the $ns^2$ $^1S_0$ - $nsnp$ $^3P^{\rm o}_0$ optical transition \cite{Mitroy-JPB-2010,Derevianko-RMP-2011,Safronova-IEEE-2012}, where the upper state is one of the metastable triplet states $^3P^{\rm o}_J$ with $J$=0, 1, and 2. Two such examples are the In$^+$ and Sr optical clocks using the 5s$^2$ $^1S_0$ - 5s5p $^3P^{\rm o}_0$ transition \cite{Safronova-PRL-2011,Ye-NC-2015}. The polarizabilities of such systems have been targeted with increasing theoretical efforts to perform precision calculations. For example, Sahoo et al. have calculated the ground-state dipole polarizability of Sr by using relativistic coupled cluster (CC) method \cite{Sahoo-PRA-2008}. Safronova et al. have given high accuracy data for the dipole polarizabilities of $5s^2$ $^1S_0$ and $5s5p$ $^3P^{\rm o}_0$ of In$^+$ and Sr using the configuration interaction (CI) +all-order method \cite{Safronova-PRL-2011,Safronova-PRA-2013-Sr}. Mitroy et al. have calculated the $5s^2$ $^1S_0$ and $5s5p$ $^3P^{\rm o}_1$ states of Sr by using CI with a semiempirical core polarization potential \cite{Mitroy-MP-2010,Cheng-PRA-2013}. Porsev et al. have calculated the Sr $5s^2$ $^1S_0$ and $5s5p$ $^3P^{\rm o}_{0,1}$ states using the CI method with many-body perturbation theory \cite{Porsev-PRA-2008}. Very recently they have calculated the $5s5p$ $^3P^{\rm o}_{1}$ state by applying the more accurate CI + all-order method \cite{Porsev-PRA-2014}. From these works, one can see that the most currently available dipole polarizability data for In$^+$ and Sr are concentrated on the $^3P^{\rm o}_0$ component, whereas the data for the $^3P^{\rm o}_1$ and $^3P^{\rm o}_2$ components remain scarce. Besides, the quadrupole moment of Sr in the $5s5p$ $^3P^{\rm o}$ state and its quadrupole polarizabilities for the $5s^2$ $^1S$ and $5s5p$ $^3P^{\rm o}$ states have been calculated in earlier works by Mitroy, et al. using the CICP method \cite{Mitroy-MP-2010, Mitroy-PRA-2004} and Porsev et al. using CI+MBPT and CI+all-order methods \cite{Derevianko-PRL-2003, Porsev-PRA-2014}. However, the recommended data for the quadrupole moments and quadrupole polarizabilities of In$^+$ remain not available.

Among various theoretical approaches in calculating polarizabilities, the finite-field method can provide reliable data if the field-dependent energies are evaluated to a high precision. This method is often implemented in computational codes for atomic and molecular property calculations~\cite{Thakkar-JCP-1988,Fleig-PRA-2005,Thiefelder-PRA-2008}. In this method the employed external field breaks the degeneracy of multiple states and thus the $J$ and $M_J$-resolved polarizabilities of the $nsnp$ $^3P^{\rm o}_J$ metastable states can be obtained directly without using basic vector algebra, where $J$ and $M_J$ are the total angular momentum and its magnetic quantum number \cite{Angel-PRPSA-1968}. In particular, the influence of the spin-orbit coupling interaction on the $J$-resolved polarizability can be revealed through comparative studies between full and scalar relativistic calculations. Successful application of the finite-field method to atoms and ions of optical frequency standard has been demonstrated for Al$^{+}$ \cite{Kallay-PRA-2011, ymyu-PRA-2013}. However, the finite-field method has not been widely applied to calculate the polarizabilities of heavy atoms and ions, for which relativistic effects and electron correlations are expected to be significant. It remains  challenging in treating these effects in this finite-field approach.

In this work, the dipole polarizabilities, dipole hyperpolarizabilities, quadrupole moments, and quadrupole polarizabilities for the ground state 5s$^2$ $^1S_0$ and the low-lying excited states 5s5p $^3P^{\rm o}_{0,1,2}$ of In$^+$ and Sr are calculated by applying the finite-field method. The convergent hierarchies of electron correlation and the convergent basis sets are adopted in the relativistic CC and CI calculations in order to obtain properties of high accuracy. The dipole polarizabilities of the ground-state 5s$^2$ $^1S_0$ and the excited state 5s5p $^3P^{\rm o}_0$, as well as the scalar and tensor polarizabilities of the excited states 5s5p $^3P^{\rm o}_{1,2}$, are obtained for In$^+$ and Sr. Comparisons with previously reported data for 5s$^2$ $^1S_0$ and 5s5p $^3P^{\rm o}_{0}$ show good agreement. The effect of the spin-orbit coupling interaction on the studied property is analyzed through comparing the $J$ and $L$-resolved values of the $5s5p$ $^3P^{\rm o}_{0,1,2}$ states, where $L$ is the total orbital angular momentum. Besides, we recommend the values of the dipole hyperpolarizabilities, the quadrupole moments, and the quadrupole polarizabilities of In$^{+}$ 5s$^2$ $^1S_0$ and 5s5p $^3P^{\rm o}_{0,1,2}$ states and Sr 5s$^2$ $^1S_0$ and 5s5p $^3P^{\rm o}_{0}$ states, for which the previous available data are rather scarce. Finally, the BBR shifts for the dipole polarizability, hyperpolarizability, and quadrupole polarizability are evaluated for the clock transition 5s$^2$ $^1S_0$ - 5s5p $^3P^{\rm o}_0$ of In$^{+}$ and Sr.

\section{Theory}
The energy shift of an atom or ion in a homogeneous electric field can be expressed as \cite{Archibong-PRA-1991}
\begin{equation}
\Delta E_{d}(F_z)=-\alpha F_z^2/2-\gamma F_{z}^4/24-\cdots,
\label{eq:t1}
\end{equation}
where $F_z$ is the electric field strength along the $z$ direction, and $\alpha$ and $\gamma$ are, respectively, the dipole polarizability and dipole hyperpolarizability. The $J$-resolved scalar and tensor polarizabilities are given by
\begin{equation}
\bar{Q}^{J}=\frac{1}{2J+1}\underset{M_{J}}{\Sigma}Q(J,M_{J})
\label{eq:t2}
\end{equation}
\begin{eqnarray}
Q_a^{J}&=Q(J,|M_{J}|=J)-\bar{Q}^{J},
\label{eq:t3}
\end{eqnarray}
where $Q$ denotes either $\alpha$ or $\gamma$, and $Q(J,M_{J})$ denotes either $\alpha$ or $\gamma$ for each $M_J$ component with $M_J$ being the projection of the angular momentum $J$ in the $z$ direction. The $L$-resolved scalar and tensor polarizabilities are also defined
by Eqs.~(\ref{eq:t2}) and (\ref{eq:t3})
with $J$ and $M_J$ replaced by $L$ and $M_L$ respectively. If the spin-orbit coupling interaction is not strong so that it can be treated as a perturbation, the relationship between the $J$- and $L$-resolved polarizabilities can be established by using the $LS$ coupling approximation~\cite{Angel-PRPSA-1968}
\begin{eqnarray}
\bar{\alpha}^J&=&\bar{\alpha}^L,  \nonumber\\
\alpha_a^J(^3P^{\rm o}_1)&=&-\alpha_a^J(^3P^{\rm o}_2)/2 = -\alpha_a^L/2.
\label{eq:t4}
\end{eqnarray}
The actual relationship between the $J$- and $L$-resolved polarizabilities can deviate from Eq.~(\ref{eq:t4}) when the spin-orbit interaction is strong and the $LS$ coupling scheme fails to describe the polarizabilities.

In a pure quadrupole electric field, the corresponding energy shift is
\begin{equation}
\Delta E_{d}(F_{zz})=-\theta F_{zz}/2-\alpha_2 F_{zz}^2/8-\cdots,
\label{eq:t5}
\end{equation}
where $F_{zz}$ is the electric field gradient in the $z$ direction, and $\theta$ and $\alpha_{2}$ are the quadrupole moment and quadrupole polarizability, respectively. The quadrupole moment $\theta$ is calculated under the condition $|M_J|=J$ or $|M_L|=L$. For the $5s5p$ $^3P$ state of In$^+$ and Sr, the $L$-resolved quadrupole moment is given for the state of $^3P^{\rm o}$ with $M_L=1$ , and the $J$-resolved quadrupole moment is given for the state of $^3P^{\rm o}_2$ with $M_J=2$.

\section{Method of Calculation}
The electric field-dependent energy is calculated at different levels of theories. Dirac-Hartree-Fock (DHF) calculations are performed that generate the reference states as well as optimized atomic orbitals. These calculations are implemented by the SCF module in the DIRAC package~\cite{Dirac}. Both the Dirac-Coulomb Hamiltonian and the Dyall Hamiltonian \cite{Dyall-ham} are employed. The Dirac-Coulomb Hamiltonian gives a complete description of the relativistic effects. We carry out the relativistic CC and CI calculations that are built on the Dirac-Coulomb Hamiltonian and yield fully relativistic results. The relativistic CC calculations are implemented by applying the MRCC suite \cite{MRCC}, and the relativistic CI calculations are implemented by applying the the KRCI module in the DIRAC package. The Dyall Hamiltonian contains only the spin-free terms with the spin-orbit interaction neglected. We carry out the scalar relativistic CI calculations that are built on the Dyall Hamiltonian and yield the scalar relativistic results. The scalar relativistic CI calculations are implemented by applying the LUCITA module in the DIRAC package.

Due to escalating computational time for adopting higher level correlation methods, the generalized active spaces (GAS) technique \cite{Knecht-JCP-2010} is used to restrict the number of correlated electrons and atomic orbitals, which makes the computation tractable. In the GAS, the Dirac-Fock orbitals are divided into the inner-core, outer-core, valence, and virtual shells. No excitation is allowed in the inner-core shells, but the excitation in the outer-core and valence shells can be defined to any order. In order to obtain convergent results, we choose the outer-core that includes increasing number of core electrons. In particular, the ${4d}$, ${4s4p4d}$, and ${3d4s4p4d}$ shells that consist of the `(core10)', `(core18)', and `core(28)' outer-cores, respectively, are considered for In$^+$. Similarly the ${4s4p}$, ${3d4s4p}$, and ${3s3p3d4s4p}$ shells that consist of the `(core8)', `(core18)', and `(core26)' outer-cores, respectively, are considered for Sr. For both In$^+$ and Sr, the valence shells are comprised of $5s5p$. The virtual orbitals with the orbital energy larger than a given cut-off value are neglected in the correlation calculations. We have carried out the calculations with the different cut-off values in order to check the convergence of the obtained results with the truncation of the virtual orbitals.

In the following, a CC calculation is referred to as `(core $n$)SD' and `(core $n$)SDT' that include single and double (SD) excitations and single, double, and triple (SDT) excitations from the outer-core and valence shells into the virtual orbitals, where $n$ means the number of outer-core electrons that are involved in electronic correlation calculations. A CI calculation is referred to as `(core $n$)SD(2in4)SDT' and `(core $n$)SD(2in4)SDTQ', where the outer-core shells are restricted to single and double (SD) excitations, and the valence shells, where two electrons are distributed in the four $5s5p$ orbitals, denoted by `(2in4)', add excited electrons from the outer core, are restricted to SDT and single, double, triple, and quadruple (SDTQ) excitations into the virtual orbitals.

The Dyall's uncontracted correlation-consistent double-, triple-, and quadruple-$\zeta$ basis sets are used, which are called $X\zeta$ with $X$=2, 3, and 4, respectively \cite{Basis-In, Basis-Sr}. Each shell is augmented by two additional diffused functions. The exponential coefficients of the augmented functions are determined according to
\begin{equation}
\zeta_{N+1}=\bigg[\frac{\zeta_N}{\zeta_{N-1}}\bigg]\zeta_{N}\,,
\label{eq:t6}
\end{equation}
where $\zeta_{N}$ and $\zeta_{N-1}$ are the two most diffused exponents for the respective atomic shells in the original basis set. Arbitrary 4-8 finite field strengths are chosen in the range of $F_z$=(0, 4.5$\times10^{-3}$) and of $F_{zz}$=(0, 4.5$\times10^{-5}$) in atomic units. The fitting is checked to remove the dependence of the properties studied on sampling. In our calculations, the criterion for energy convergence is set to be $10^{-10} $ Hartree.

We use the composite scheme \cite{Kallay-PRA-2011} to give the final value $P_{\rm Final}$ of a studied property. In the CC calculation,
\begin{equation}
P_{\rm Final}=P_{SD}+\Delta P_{T}+\Delta P_{\rm core},
\label{eq:t7}
\end{equation}
where $P_{SD}$ is the value calculated for property $P$ using the (core10)SD method with the $X$=$4\zeta$ basis set, $\Delta P_{T}$ and $\Delta P_{\rm core}$ are the corrections due to the triple excitation and more outer-core electrons. In the CI calculation,
\begin{equation}
P_{\rm Final}=P_{SDT}+\Delta P_{Q}+\Delta P_{\rm core},
\label{eq:t8}
\end{equation}
where $P_{SDT}$ is the value calculated for property $P$ using the (core10)SD(2in4)SDT method with the $X$=$4\zeta$ basis set, and $\Delta P_{Q}$ is the correction due to the quadruple excitation. The virtual orbitals with the orbital energy larger than 20 a.u. for In$^{+}$ and 10 a.u. for Sr are neglected in the correlation calculations. We also carry out the corresponding calculations of the cut-off value of the virtual orbitals of 100 a.u. for In$^{+}$ and Sr. The truncation of virtual orbitals of the cut-off value 100 a.u. results in change of $\alpha^J$ less than 0.5\% in comparison with the cases of the cut-off value 20 a.u. for In$^{+}$ and 10 a.u. for Sr. The contribution of the virtual orbitals with the orbital energy larger than 20 a.u. for In$^{+}$ and 10 a.u. for Sr is therefore omitted in the composite scheme.

The uncertainty in $P_{\rm Final}$ is mainly caused by three possible error sources. The first error is due to the finite basis set used for calculating $P_{SD}$ and $P_{SDT}$. The convergence of $P$ with respect to the basis set is very quick when the basis set is larger than $X$=$3\zeta$, as shown by the Al$^{+}$ results \cite{ymyu-PRA-2013}. Thus, we assume empirically that the error for $P_{SD}$ and for $P_{SDT}$ is equal to half of the difference of the $P$ values calculated by using $X$=4$\zeta$ and $3\zeta$ basis sets. This error considers the possible correction with respect to the value computed in the infinite basis. The second error comes from the estimation of $\Delta P_{T}$ and $\Delta P_{Q}$. Previous experience \cite{ymyu-PRA-2013} has shown that the $\Delta P_{T}$ and $\Delta P_{Q}$ contributions computed even with the smaller 2$\zeta$ basis set are never in error by more than 50\% with respect to the basis-set limit, and hence the error bar of $\Delta P_{T}$ and of $\Delta P_{Q}$ is taken to be half of itself, i.e., $0.5\times\Delta P_{T}$ and $0.5\times\Delta P_{Q}$. The third error is due to the estimation of $\Delta P_{\rm core}$. As we will show in the following, the studied properties start to converge even with medium-size outer cores such as core(18), which indicates that the error of $\Delta P_{\rm core}$ is not larger than $P_{(\rm core28)}-P_{(\rm core18)}$ for In$^+$ and $P_{(\rm core26)}-P_{(\rm core18)}$ for Sr. Hence, the overall uncertainty in $P_{\rm Final}$ can be estimated to be Root-Mean-Square(RMS) of such three errors.

Throughout this paper, atomic units (a.u.) are used, unless otherwise stated. The atomic units of $\alpha$, $\alpha_2$, $\gamma$, $F_z$, and $F_{zz}$ are, respectively, $1.648778\times10^{-41}$C$^2$m$^2$J$^{-1}$, $4.617048\times10^{-62}$C$^2$m$^4$J$^{-1}$, $6.235378\times10^{-65}$C$^4$m$^4$J$^{-3}$, 5.142250$\times$10$^{9}$ V/cm, and 5.142250$\times$10$^{11}$ V/cm$^2$.

\begin{table*}[btp]
\renewcommand{\arraystretch}{1.2}
\setlength{\tabcolsep}{3pt}
\scriptsize
\caption{Dipole polarizabilities $\alpha^J$ of In$^{+}$ for the states of $5s^2$ $^1S_0$ and $5s5p$ $^3P^{\rm o}_{0,1,2}$ calculated by the relativistic CC method, where (core10), (core18), and (core28) correspond respectively to the $4d$, $4s4p4d$, and $3d4s4p4d$ core shells included in the electron correlation calculations.}
\label{table1}
\begin{tabular}{p{3.6cm}    p{1.0cm}    p{1.0cm}       p{1.0cm}            p{1.0cm} p{1.0cm} p{1.0cm} p{1.0cm} p{1.0cm}}\hline\hline
 \multirow{2}{*}{Level of excitation $^a$}&\multirow{2}{*}{$^1S_{0}$} &\multirow{2}{*}{$^3P^{\rm o}_{0}$}&$^3P^{\rm o}_{1}$ &\multicolumn{5}{c}{$^3P^{\rm o}_{2}$}    \\ \cline{5-9}
      &        &                   & $|M_J|$=1             & $|M_J|$=0& $|M_J|$=1& $|M_J|$=2&$\bar{\alpha}^J$&$\alpha_a^J$  \\\hline
 Basis:2$\zeta$(23s,17p,13d,4f)      & \multicolumn{7}{c}{ }     \\
 (core10)SD         &24.83      &27.91                   &29.12                           &32.34   &31.47   &28.88   &30.61  &--1.73  \\
 (core10)SDT        &24.54      &27.50                   &28.75                           &32.26   &31.13   &28.42   &30.27  &--1.85 \\
 $\Delta P_T$              &--0.29    &--0.41     &--0.37    &--0.08    &--0.35    &--0.46    &--0.34    &--0.12   \\
 Error in $\Delta P_T$     &$\pm$0.15&$\pm$0.21 &$\pm$0.19&$\pm$0.04&$\pm$0.18&$\pm$0.23&$\pm$0.17&$\pm$0.06 \\
 Basis:3$\zeta$(30s,23p,17d,5f,3g)   & \multicolumn{7}{c}{ }     \\
 (core10)SD         &24.90      &27.01                   &27.98                           &30.75   &30.09   &28.10   &29.42  &--1.33  \\
 (core18)SD         &24.71      &26.81                   &27.83                           &30.68   &30.09   &27.84   &29.31  &--1.47 \\
 (core28)SD         &24.67      &26.77                   &27.79                           &30.63   &29.93   &29.79   &29.21  &--1.42 \\
 $\Delta P_{\rm core}$         &--0.24    &--0.25     &--0.19    &--0.12    &--0.16    &--0.30    &--0.21    &--0.09  \\
 Error in $\Delta P_{\rm core}$&$\pm$0.04&$\pm$0.04&$\pm$0.04 &$\pm$0.05&$\pm$0.16&$\pm$0.05&$\pm$0.09&$\pm$0.05 \\
 Basis:4$\zeta$(35s,29p,20d,7f,5g,3h)& \multicolumn{7}{c}{ }     \\
(core10)SD,$P_{SD}$      &24.86    &26.91     &27.87    &30.64    &29.98    &28.01    &29.32    &--1.31  \\
Error in $P_{SD}$        &$\pm$0.03&$\pm$0.05 &$\pm$0.06&$\pm$0.06&$\pm$0.05&$\pm$0.04&$\pm$0.05&$\pm$0.01 \\
\multicolumn{9}{c}{ $P_{\rm Final}$=$P_{SD}$+$\Delta P_{\rm core}$+$\Delta P_T$}              \\
Final data, $P_{\rm Final}$    &24.33    &26.25     &27.31    &30.44    &29.48    &27.25    &28.78    &--1.53 \\
Uncertainty($\%$)          &0.62     &0.82      &0.73     &0.28     &0.82     &0.87     &0.69     &4.99 \\\hline
Ref.\cite{Safronova-PRL-2011}&24.01  &26.02     &         &         &         &         &         &      \\ \hline \hline
\multicolumn{9}{l}{$^a$ The relativistic CI calculation is performed using the $3\zeta$ basis set at (core10)SD(2in4)SD$<$2 level, which yields} \\
\multicolumn{9}{l}{ $\alpha$=25.06 and 27.93 for $^1S_{0}$ and $^3P^{\rm o}_{0}$, $\bar{\alpha}^J$=28.59 and $\alpha_a^J$=0.34 for $^3P^{\rm o}_1$, and
$\bar{\alpha}^J$=30.30 and $\alpha_a^J$=--1.38 for $^3P^{\rm o}_2$.  } \\
\end{tabular}
\end{table*}

\begin{table*}[btp]
\renewcommand{\arraystretch}{1.2}
\setlength{\tabcolsep}{3pt}
\scriptsize
\caption{Dipole hyperpolarizabilities $\gamma^J$ of In$^{+}$ for the states of $5s^2$ $^1S_0$ and $5s5p$ $^3P^{\rm o}_{0,1,2}$  calculated by the relativistic CC method, where (core10), (core18), and (core28) correspond respectively to the $4d$, $4s4p4d$, and $3d4s4p4d$ core shells included in the electron correlation calculations.}
\label{table2}
\begin{tabular}{p{3.6cm}    p{1.0cm}    p{1.0cm}       p{1.0cm}            p{1.0cm} p{1.0cm} p{1.0cm} p{1.0cm} p{1.0cm}}\hline\hline
 \multirow{2}{*}{Level of excitation $^a$}&\multirow{2}{*}{$^1S_{0}$} &\multirow{2}{*}{$^3P^{\rm o}_{0}$}&$^3P^{\rm o}_{1}$ &\multicolumn{5}{c}{$^3P^{\rm o}_{2}$}    \\ \cline{5-9}
      &        &                   & $|M_J|$=1             & $|M_J|$=0& $|M_J|$=1& $|M_J|$=2&$\bar{\gamma}^J$&$\gamma_a^J$  \\\hline
 Basis:2$\zeta$(23s,17p,13d,4f)    & \multicolumn{8}{c}{ }     \\
 (core10)SD         &3695           &14752            &21119                   &27116  &20729   &7294    &16632         &--9338  \\
 (core10)SDT        &3640           &15134            &21611                   &27546  &20989   &7773    &17014         &--9241 \\
 $\Delta P_T$	    &--55            &381              &492                   &431     & 260    &479     &382         &97 \\
 Error in $\Delta P_T$&$\pm$28       &$\pm$191         &$\pm$246              &$\pm$216&$\pm$130&$\pm$240&$\pm$191    &$\pm$49 \\
 Basis:3$\zeta$(30s,23p,17d,5f,3g)  & \multicolumn{8}{c}{ }     \\
 (core10)SD         &3143           &13435            &19265                   &26644  &20951   &6820    &16437           &--9617  \\
 (core18)SD         &3072           &13467            &19368                   &26472  &20755   &6968    &16384           &--9645 \\
 (core28)SD         &3065           &13464            &19327                   &26480  &20703   &7056    &16400           &--9344 \\
 $\Delta P_{\rm core}$	&--78            &28               &62                    &--164    & --248   &236     &--38         &273 \\
 Error in $\Delta P_{\rm core}$&$\pm$7 &$\pm$3          &$\pm$42               &$\pm$9&$\pm$52&$\pm$88 &$\pm$16     &$\pm$72 \\
 Basis:4$\zeta$(35s,29p,20d,7f,5g,3h)& \multicolumn{8}{c}{ }     \\
 (core10)SD,$P_{SD}$&3122           &13057            &19125                 &26265   &20571   &6765    &16187       &--9422 \\
 Error in $P_{SD}$  &$\pm$11        &$\pm$189         &$\pm$70               &$\pm$190&$\pm$190&$\pm$28 &$\pm$125    &$\pm$97 \\
\multicolumn{9}{c}{  $P_{\rm Final}$=$P_{SD}$+$\Delta P_{\rm core}$+$\Delta P_T$ }              \\
 Final data, $P_{\rm Final}$&2989          &13467            &19679                 &26532   &20583   &7479    &16531       &--9052 \\
 Uncertainty($\%$)   &1.01          &1.99             &1.32                  &1.08    &1.15    &3.43    &1.38        &1.44  \\ \hline \hline
\multicolumn{9}{l}{$^a$ The relativistic CI calculation is performed using the $3\zeta$ basis set at (core10)SD(2in4)SD$<$2 level, which yields } \\
\multicolumn{9}{l}{ $\alpha$=2715 and 16164 for $^1S_{0}$ and $^3P^{\rm o}_{0}$, $\bar{\alpha}^J$=17247 and $\alpha_a^J$=4263 for $^3P^{\rm o}_1$, and
$\bar{\alpha}^J$=18614 and $\alpha_a^J$=--8518 for $^3P^{\rm o}_2$.  } \\\end{tabular}
\end{table*}

\begin{table*}[btp]
\renewcommand{\arraystretch}{1.2}
\setlength{\tabcolsep}{3pt}
\scriptsize
\caption{Dipole polarizabilities $\alpha^L$ and hyperpolarizabilities $\gamma^L$ for the $5s^2$ $^1S$ and $5s5p$ $^3P^{\rm o}$ states of In$^{+}$ obtained by the scalar relativistic CI method, where (core10), (core18), and (core28) correspond respectively to the $4d$, $4s4p4d$, and $3d4s4p4d$ core shells included in the electron correlation calculations.}
\label{table3}
\begin{tabular}{p{3.6cm}        p{1.0cm}      p{1.0cm}   p{1.0cm}  p{1.0cm}  p{1.0cm} p{0.2cm} p{1.0cm}  p{1.0cm}  p{1.0cm}  p{1.0cm}   p{1.0cm}     }\hline\hline
 \multirow{3}{*}{Level of excitation}& \multicolumn{5}{c}{$\alpha$}                          & & \multicolumn{5}{c}{$\gamma$} \\ \cline{2-6} \cline{8-12}
                                     & \multirow{2}{*}{$^1S_{0}$}&\multicolumn{4}{c}{$^3P^{\rm o}$}& &\multirow{2}{*}{$^1S_{0}$}& \multicolumn{4}{c}{$^3P^{\rm o}$}  \\ \cline{3-6} \cline{9-12}
                                     &       &$|M_L|$=0&$|M_L|$=1&$\bar{\alpha}^L$&$\alpha_a^L$  & &         &$|M_L|$=0&$|M_L|$=1 & $\bar{\gamma}^L$&$\gamma_a^L$           \\\hline
 Basis:2$\zeta$(23s,17p,13d,4f)      & \multicolumn{11}{c}{ }     \\
 (core10)SD(2in4)SDT           &24.42   &32.24  &27.98   &29.40    &--1.42   & &   2832  &35528     &7099     &16576     &--9476  \\
 (core18)SD(2in4)SDT           &24.28   &32.23  &27.84   &29.31    &--1.46   & &   2521  &35712     &6972     &16552     &--9580  \\
 (core28)SD(2in4)SDT           &24.26   &32.19  &27.82   &29.28    &--1.46   & &   2567  &35679     &6967     &16538     &--9571  \\
 (core10)SDT(2in4)SDTQ         &24.41   &32.17  &27.82   &29.27    &--1.45   & &   3650  &35577     &7381     &16780     &--9398  \\
 $\Delta P_{Q}$	               &--0.015  &--0.07     &--0.16    &--0.13     &--0.03     & &  789   &49      &282     &204     &78   \\
 Error in $P_{Q}$              &$\pm$0.008&$\pm$0.035&$\pm$0.08&$\pm$0.065&0.015     & &$\pm$395&$\pm$25 &$\pm$141&$\pm$102&$\pm$39    \\
 $\Delta P_{\rm core}$	   &--0.16   &--0.05     &--0.16    &--0.12     &--0.04     & &  --294  &151     &--132    &--38     &--94   \\
 Error in $P_{\rm core}$   &$\pm$0.02&$\pm$0.05 &$\pm$0.0.02&$\pm$0.03&$\pm$0.01& &$\pm$46&$\pm$33&$\pm$5&$\pm$14 &$\pm$9    \\
Basis:3$\zeta$(30s,23p,17d,5f,3g)   & \multicolumn{11}{c}{ }     \\
 (core10)SD(2in4)SDT           &24.38    &30.43  &27.17   &28.26    &--1.08   & &   2182  &34283     &6742   &15922     &--9180  \\
 Basis:4$\zeta$(35s,29p,20d,7f,5g,3h)& \multicolumn{11}{c}{ }     \\
 (core10)SD(2in4)SDT, $P_{SDT}$&24.34    &30.35     &27.12    &28.20     &--1.08     & &2270    &34257   &6761    &15926   &--9165  \\
 Error in $P_{SDT}$	           &$\pm$0.02&$\pm$0.04 &$\pm$0.03&$\pm$0.03 &$\pm$0.005& &$\pm$44 &$\pm$13 &$\pm$9  &$\pm$2  &$\pm$8    \\
& \multicolumn{11}{c}{  $P_{\rm Final}$=$P_{SDT}$+$\Delta P_{\rm core}$+$\Delta P_{Q}$}             \\
 Final data, $P_{\rm Final}$       &24.16    &30.22     &26.80    &27.94     &--1.14     & &   2765 &34457   &6910    &16092   &--9182 \\
 Uncertainty($\%$)             &0.13     &0.23      &0.33     &0.28      &1.56      & &   14.46&0.13    &2.05    &0.64    &0.44 \\ \hline \hline
\end{tabular}
\end{table*}

\section{Results and discussion}

\subsection{Dipole polarizability and hyperpolarizability}

Table~\ref{table1} summarizes the results of $\alpha^J$ for the $5s^2$ $^1S_0$ and $5s5p$ $^3P^{\rm o}_{0,1,2}$ states of In$^+$ calculated by using the relativistic CC method. Firstly, the obtained values of $\alpha^J$ in the (core10)SD calculation with the $X$=$2\zeta$, $3\zeta$, and $4\zeta$ basis sets show a good convergence. The error in $P_{SD}$ is only about 0.01$\sim$0.06, which implies that the correction with respect to the infinite basis set is very small. Then, the $\Delta P_{\rm core}$ correction is estimated with the $X$=$3\zeta$ basis set. Upon inclusion of the $4s4p4d$ core electrons into the (core18)SD calculation, $\alpha^J$ decreases in comparison with the (core10)SD case. The effect of adding more core electrons, $3d4s4p4d$, is illustrated by the rather small difference between the (core28)SD and (core18)SD calculations, indicating that $\alpha^J$ has entered into convergence region. Thus, $\Delta P_{\rm core}$ is estimated as the difference between the (core28)SD and (core10)SD calculations, which is about 0.09$\sim$0.30. By comparison of the (core10)SD, (core18)SD, and (core28)SD calculations, we see that the minimum number of core electrons that needs to be correlated for In$^+$ is 18; in other words, the $4s4p4d$ core electrons need to be correlated in order to obtain accurate $\alpha^J$. Next, $\Delta P_{T}$ is estimated by the difference between the (core10)SD and (core10)SDT values with the $X=2\zeta$ basis set, which is around $-0.08\sim-0.46$. Both the $\Delta P_{T}$ and $\Delta P_{\rm core}$ corrections are large in comparison with the error in $P_{SD}$, and thus an omission of such a correction would lead to an underestimation of $\alpha^J$. The reference data for the $5s^2$ $^1S_0$ and $5s5p$ $^3P^{\rm o}_0$ states of In$^{+}$ are also given in Table~\ref{table1}, as calculated by Safronova et al. by using the CI+all-order method~\cite{Safronova-PRL-2011}. Our results are consistent with their data with a discrepancy around 1\%. Finally, we recommend that $\bar{\alpha}^J$=27.31 for $5s5p$ $^3P^{\rm o}_1$ with $|M_J|$=1 and $\bar{\alpha}^J$=28.78 and $\alpha_a^J=-1.53$ for $5s5p$ $^3P^{\rm o}_2$.

Our calculated results for $\gamma^J$ are listed in Table~\ref{table2}. The $\gamma^J$ values show a similar convergence trend with respect to the size of basis set. The values of $\Delta P_{T}$, and $\Delta P_{\rm core}$ are also larger than the error in $P_{SD}$ for $\gamma^J$, similar to the case of $\alpha^J$. As shown in Tables~\ref{table1} and \ref{table2}, the largest source of uncertainty in the final data comes from the error in $\Delta P_{T}$ for both $\alpha^J$ and $\gamma^J$. As mentioned above, $\Delta P_{T}$ has important contribution to the final data. In this work, due to very high computational demand in using larger basis sets, $\Delta P_{T}$ is calculated only with $X=2\zeta$. This basis set is much smaller, which may cause an overestimation of the uncertainty of $\Delta P_{T}$, like the cases of $\alpha^J_a$ for $5s5p$ $^3P^{\rm o}_2$ and $\gamma^J$ for $5s5p$ $^3P^{\rm o}_2$ with $M_J=2$. Using the same composite scheme of convergence, we arrive at the recommended values of $\gamma^J$=2989, 13467, and 19679 for the $5s^2$ $^1S_0$, $5s5p$ $^3P^{\rm o}_0$, and $5s5p$ $^3P^{\rm o}_1$, $|M_J|$=1 states, respectively, and $\bar{\gamma}^J$=16531 and $\gamma_a^J=-9052$ for the $5s5p$ $^3P^{\rm o}_2$ state. The uncertainties of the final values of $\gamma$ are slightly larger than those for $\alpha^J$, which is comprehensible because the hyperpolarizability is a higher-order response that is more sensitive to a small energy variation and thus various contributions bring substantial corrections to $\gamma^J$. The relativistic CI calculation is implemented at the level of (core10)SD(2in4)SD$<$2, where the cutoff for the virtual orbitals is 2 a.u. in energy, and the basis set is $3\zeta$, as indicated at the footnotes of Tables~\ref{table1} and \ref{table2}.

The results of $\alpha^L$ and $\gamma^L$ for the $5s^2$ $^1S$ and $5s5p$ $^3P^{\rm o}$ states of In$^{+}$, calculated by the scalar relativistic CI method, are contained in Table~\ref{table3}. $P_{SDT}$ is determined using the $X$=$4\zeta$ basis set, and $\Delta P_{Q}$ and $\Delta P_{\rm core}$ are determined using the $X$=$2\zeta$ basis set as the differences between the (core10)SD(2in4)SDT and (core10)SDT(2in4)SDTQ, and between the (core10)SD(2in4)SDT and (core28)SD(2in4)SDT, respectively. The triple excitation is considered in the scalar relativistic CI calculations, which reduces the uncertainties in the final values of $\alpha^L$ and $\gamma^L$, as compared with the relativistic CC calculations. In Table~\ref{table3}, the $\Delta P_{Q}$ and $\Delta P_{\rm core}$ corrections are also larger than the error in $P_{SDT}$. This trend is in accordance with that in the relativistic CC calculations. Because $\Delta P_{Q}$ is computed only with a much smaller basis set, an estimation of error $\Delta P_{Q}$ can not be exact (like $\Delta P_T$ in Tables~\ref{table1} and \ref{table2}), which leads to an anomalously large uncertainty in the final data for the $5s^2$ $^1S_0$ state. Up to now, we can find that, for In$^+$, increasing basis set up to $X=4\zeta$ has decreased the error in $P_{SD}$ with respect to the infinite basis set. However, the $\Delta P_T$ and $\Delta P_{\rm core}$ corrections are larger than the error in $P_{SD}$. In this situation, both $\Delta P_T$ and $\Delta P_{\rm core}$ become crucial for an accurate evaluation of the dipole polarizability and hyperpolarizability of In$^+$.

\begin{table*}[btp]
\renewcommand{\arraystretch}{1.2}
\setlength{\tabcolsep}{3pt}
\scriptsize
\caption{Dipole polarizabilities $\alpha^J$ for the states of $5s^2$ $^1S_0$ and $5s5p$ $^3P^{\rm o}_{0,1,2}$ of Sr calculated by the relativistic CC method, where (core8), (core18), and (core26) correspond respectively to the $4s4p$, $3d4s4p$, and $3s3p3d4s4p$ core shells included in the electron correlation calculations.}
\label{table4}
\begin{tabular}{p{3.6cm}                  p{1.0cm}   p{1.0cm}   p{1.80cm}    p{1.0cm}  p{1.0cm}  p{1.0cm} p{1.0cm}  p{0.7cm}}\hline\hline
 \multirow{2}{*}{Level of excitation $^{a}$}&\multirow{2}{*}{$^1S_{0}$}&\multirow{2}{*}{$^3P^{\rm o}_{0}$}&$^3P^{\rm o}_{1}$ &\multicolumn{5}{c}{$^3P^{\rm o}_{2}$} \\ \cline{5-9}
                                          &          &          & $|M_J|$=1 &$|M_J|$=0&$|M_J|$=1&$|M_J|$=2&$\bar{\alpha}^J$&$\alpha_a^J$  \\\hline
 Basis:2$\zeta$(23s,17p,12d,3f)           & \multicolumn{8}{c}{ }     \\
 (core8)SD                                &205.1     &349.3     &377.1      &418.7    &397.1    &331.7    &375.3    &--43.5    \\
 (core8)SDT                               &200.1     &348.8     &379.5      &422.2    &499.8    &332.0    &377.2    &--45.2    \\
 $\Delta P_T$	                          &--5.0      &--0.5      &2.4        &3.5      &2.8      &0.2      &1.9      &--1.7     \\
 Error in $\Delta P_T$                    &$\pm$2.5  &$\pm$0.3  &$\pm$1.2   &$\pm$1.8 &$\pm$1.4 &$\pm$0.1 &$\pm$1.0 &$\pm$0.9 \\
 Basis:3$\zeta$(31s,22p,15d,7f,3g)        & \multicolumn{8}{c}{ }     \\
 (core8)SD                                &204.8     &423.8     &460.8      &515.1    &490.5    &415.7    &465.5    &--49.8    \\
 (core18)SD                               &204.6     &415.3     &450.3      &503.8    &478.8    &406.3    &454.8    &--48.5    \\
 (core26)SD                               &204.7     &418.8     &454.4      &509.0    &486.9    &410.1    &460.6    &--50.5    \\
 $\Delta P_{\rm core}$	                      &--0.2      &--5.0   &--6.4       &--6.2   &--7.6     &--5.6     &--6.5     &0.9      \\
 Error in $\Delta P_{\rm core}$	              &$\pm$0.01 &$\pm$3.5  &$\pm$4.0   &$\pm$5.2 &$\pm$4.1 &$\pm$3.9 &$\pm$4.2 &$\pm$0.3 \\
 Basis:4$\zeta$(35s,27p,17d,9f,7g,3h)     & \multicolumn{8}{c}{ }     \\
 (core8)SD, $P_{SD}$                      &204.8     &446.4     &484.9      &546.6    &520.0    &441.9    &494.1    &--52.2    \\
 Error in $P_{SD}$                        &0.0       &$\pm$11.3 &$\pm$12.1  &$\pm$15.7&$\pm$14.8&$\pm$13.1&$\pm$14.3&$\pm$1.2 \\
\multicolumn{9}{c}{$P_{\rm Final}$=$P_{SD}$+$\Delta P_{\rm core}$+$\Delta P_T$ }         \\
 Final data,$P_{\rm Final}$                   &199.7     &444.1     &480.9      &543.9    &519.2    &436.5    &491.1    &--54.5 \\
 Uncertainty($\%$)                        &1.2       &2.7       &2.7        &3.1      &3.0      &3.1      &3.1      &2.8   \\\hline
 RCC  \cite{Sahoo-PRA-2008}               &199.7     &          &                &         &         &         &         &            \\
 CI+MBPT \cite{Porsev-PRA-2008}           &197.2     &457.0     &498.8           &         &         &         &         &            \\
 CI+all-order \cite{Safronova-PRA-2013-Sr}&194.4     &441.9     &                &         &         &         &         &            \\
 CICP \cite{Mitroy-MP-2010}               &204.5     &          &497.0 (27.7)$^b$&         &         &         &         &     \\
CI+all-order  \cite{Porsev-PRA-2014}      &          &          &459.2 (26.0)$^b$&         &         &         &         &     \\
 Expt. \cite{Oppen-Exper-1969}            &          &          &(24.5)$^b$          &         &         &         &         & \\\hline \hline
\multicolumn{9}{l}{$^a$ The relativistic CI calculation is performed using the $3\zeta$ basis set at (core8)SD(2in4)SD$<$2 level, which yields }\\
\multicolumn{9}{l}{$\alpha$=179.5 and 394.5 for $5s^2$ $^1S_{0}$ and $5s5p$ $^3P^{\rm o}_{0}$, $\bar{\alpha}^J$=409.7 and $\alpha_a^J$=23 for $5s5p$ $^3P^{\rm o}_1$, and $\bar{\alpha}^J$=455.2 and $\alpha_a^J$=--51.8 } \\
\multicolumn{9}{l}{for $5s5p$ $^3P^{\rm o}_2$.}\\
\multicolumn{9}{l}{$^b$ Here given are the scalar and tensor (in parentheses) polarizabilities for $5s5p$ $^3P^{\rm o}_1$. }\\
\end{tabular}
\end{table*}

\begin{table}[btp]
\renewcommand{\arraystretch}{1.2}
\setlength{\tabcolsep}{3pt}
\scriptsize
\caption{Dipole hyperpolarizabilities $\gamma^J$ for the states of $5s^2$ $^1S_0$ and $5s5p$ $^3P^{\rm o}_{0}$ of Sr obtained by the relativistic CC method, where (core8), (core18), and (core26) correspond respectively to the $4s4p$, $3d4s4p$, and $3s3p3d4s4p$ core shells included in the electron correlation calculations.}
\label{table5}
\begin{tabular}{p{4.0cm}        p{1.5cm}      p{1.5cm}   }\hline\hline
 Level of excitation            &$^1S_0$  & $^3P^{\rm o}_0$         \\ \hline
 Basis:2$\zeta$(23s,17p,12d,3f) & \multicolumn{2}{c}{ }     \\
 (core8)SD                      &484500     &7050552             \\
 (core8)SDT                     &510144     &6905904             \\
 $\Delta P_{T}$	                &25644      &--144648             \\
 Error in $\Delta P_{T}$        &$\pm$12822 &$\pm$72324          \\
 Basis:3$\zeta$(31s,22p,15d,7f,3g)    & \multicolumn{2}{c}{ }    \\
 (core8)SD                      &680965     &3568344             \\
 (core18)SD                     &672720     &3452688             \\
 (core26)SD                     &674592     &3289040             \\
 $\Delta P_{\rm core}$	            &--6373      &--279304             \\
 Error in $\Delta P_{\rm core}$     &$\pm$1872  &$\pm$163648         \\
 Basis:4$\zeta$(35s,27p,17d,9f,7g,3h) & \multicolumn{2}{c}{ }    \\
 (core8)SD, $P_{SD}$   &672686     &3652171             \\
 Error in $P_{SD}$              &$\pm$8279  &$\pm$83827          \\
\multicolumn{3}{c}{$P_{\rm Final}$=$P_{SD}$+$\Delta P_{\rm core}$+$\Delta P_{T}$ } \\
 Final data, $P_{\rm Final}$        &691957     &3228219             \\
 Uncertainty($\%$)              &2.22       &6.12                \\ \hline\hline
\end{tabular}
\end{table}

\begin{table}[btp]
\renewcommand{\arraystretch}{1.2}
\setlength{\tabcolsep}{3pt}
\scriptsize
\caption{Dipole polarizabilities $\alpha^L$ for the states of $5s^2$ $^1S$ and $5s5p$ $^3P^{\rm o}$ of Sr obtained by the scalar relativistic CI method, where (core8), (core18), and (core26) correspond respectively to the $4s4p$, $3d4s4p$, and $3s3p3d4s4p$ core shells included in the electron correlation calculations.}
\label{table6}
\begin{tabular}{p{3.3cm}        p{0.83cm}      p{0.83cm}   p{0.83cm}  p{0.83cm}  p{0.83cm}  }\hline\hline
 \multirow{2}{*}{Level of excitation}& \multirow{2}{*}{$^1S_0$} & \multicolumn{4}{c}{$^3P^{\rm o}$}                    \\ \cline{3-6}
                                     &                          &$|M_L|$=0&$|M_L|$=1&$\bar{\alpha}^L$&$\alpha_a^L$ \\ \hline
 Basis:2$\zeta$(23s,17p,12d,3f)      & \multicolumn{5}{c}{ }     \\
 (core8)SD(2in4)SDT             &199.5   &443.7  &312.6         &356.3    &--43.7      \\
 (core18)SD(2in4)SDT            &198.2   &427.5  &302.3         &344.0    &--41.8      \\
 (core26)SD(2in4)SDT            &197.8   &427.6  &302.3         &344.1    &--41.7      \\
 (core8)SDT(2in4)SDTQ           &198.7   &457.2  &325.2         &369.3    &--44.0      \\
 $\Delta P_{Q}$	                &--0.7     &13.5     &12.6      &12.9      &--0.3        \\
 Error in $\Delta P_{Q}$        &$\pm$0.4&$\pm$6.8 &$\pm$6.3  &$\pm$6.5  &$\pm$0.2   \\
$\Delta P_{\rm core}$	            &--1.6    &--16.2     &--10.3     &--12.3     &2.0        \\
 Error in $\Delta P_{\rm core}$     &$\pm$0.35&$\pm$0.01&$\pm$0.05 &$\pm$0.04 &$\pm$0.04  \\
Basis:3$\zeta$(31s,22p,15d,7f,3g)    & \multicolumn{5}{c}{ }     \\
 (core8)SD(2in4)SDT             &198.4   &528.4  &381.9         &430.7    &--48.8      \\
 Basis:4$\zeta$(35s,27p,17d,9f,7g,3h) & \multicolumn{5}{c}{ }     \\
 (core8)SD(2in4)SDT, $P_{SD}$   &197.1   &547.7     &396.7     &447.0     &--50.3        \\
 Error in $P_{SD}$              &$\pm$0.7&$\pm$9.6  &$\pm$7.4  &$\pm$8.1  &$\pm$0.7     \\
&  \multicolumn{5}{c}{ $P_{\rm Final}$=$P_{SDT}$+$\Delta P_{\rm core}$+$\Delta P_{Q}$ }          \\
 Final data, $P_{\rm Final}$        &194.7   &544.9     &398.9     &447.6     &--48.7      \\
 Uncertainty($\%$)              &1.3     &1.8       &2.5       &2.1       &3.5        \\ \hline \hline
\end{tabular}
\end{table}

\begin{table}[btp]
\renewcommand{\arraystretch}{1.2}
\setlength{\tabcolsep}{3pt}
\scriptsize
\caption{Quadrupole moments $\theta$ and polarizabilities $\bar{\alpha}_2$, where the $L$-resolved values are obtained by the scalar relativistic CI method and the $J$-resolved values are obtained by the relativistic CC method for $5s5p$ $^3P^{\rm o}_{0,2}$ and the relativistic CI method for $5s5p$ $^3P^{\rm o}_{1}$.}
\label{table7}
\begin{tabular}{ p {1.0cm} p{1.0cm}  p{1.6cm}      p{2.5cm}   }\hline\hline
          &       &$\theta^L$                &$\theta^J$        \\\hline
In$^+$    &$^1S$&                          &                   \\
          &$^3P^{\rm o}$&4.36                      &4.64 ($^3P^{\rm o}_2$)   \\ [+1 ex]
Sr        &$^1S$&                          &                     \\
          &$^3P^{\rm o}$&15.56                     &15.76 ($^3P^{\rm o}_2$)  \\
          &       &                          &15.6 \cite{Derevianko-PRL-2003}, 15.52 \cite{Mitroy-PRA-2004}   \\[+2 ex]\hline
          &       &$\bar{\alpha}_{2}^L$      &$\bar{\alpha}_{2}^J$\\[+2 ex]
In$^+$    &$^1S$&127                       &129                 \\
          &$^3P^{\rm o}$&145                       &1425 ($^3P^{\rm o}_0$)              \\
          &       &                          &1678 ($^3P^{\rm o}_1$)              \\
          &       &                          &--859.3 ($^3P^{\rm o}_2$)            \\[+1 ex]
Sr        &$^1S$&4688                      &4608                \\
          &       &4640 \cite{Mitroy-MP-2010}&4545 \cite{Porsev-PRA-2014} \\
          &$^3P^o$&6756                      &9.75$\times$10$^4$ ($^3P^{\rm o}_0$) \\
          &       &6949 \cite{Mitroy-MP-2010}&1.17$\times$10$^5$ ($^3P^{\rm o}_1$) \\
          &       &                          &1.05$\times$10$^5$ \cite{Porsev-PRA-2014}\\
          &       &                          & --7.39$\times$10$^4$ ($^3P^{\rm o}_2$) \\ \hline\hline
\end{tabular}
\end{table}

Table~\ref{table4} summarizes the results of $\alpha^J$ for the states of $5s^2$ $^1S_0$ and $5s5p$ $^3P^{\rm o}_{0,1,2}$ of Sr, calculated by using the relativistic CC method. The relativistic CI calculations, implemented at the (core8)SD(2in4)SD$<$2 level with the $3\zeta$ basis set, are presented at the footnote area for comparison. As seen from the (core8)SD data with the $X$=2, 3, and 4$\zeta$ basis sets in Table~\ref{table4}, the changes of $\alpha^J$ for Sr $5s^2$ $^1S_0$ state are very small , while the $\alpha^J$ values for Sr $5s5p$ $^3P^o_{0,1,2}$ states tend to increase obviously for the larger basis sets, implying the basis set effect is rather large for these states. The error in $P_{SD}$ for the $5s5p$ $^3P^{\rm o}_{0,1,2}$ states of Sr is about 12.1$\sim$15.7, except for $\alpha^J_a$ of $5s5p$ $^3P^{\rm o}_{2}$, which is larger than $\Delta P_T$ and $\Delta P_{\rm core}$. This indicates that the primary dominant factor that affects the final $\alpha^J$  of Sr comes from the size of the basis set. This trend differs from the results mentioned earlier for In$^+$. Note that the error in $\Delta P_{SD}$ is very small and is far less than $\Delta P_T$ and $\Delta P_{\rm core}$ for the case of In$^+$. The possible reason for such a difference is because of the different electron distribution of In$^{+}$ and Sr. As a positively charged ion, in general In$^{+}$ has compact electron distribution around nucleus. However, as a neutral atom, Sr would have more diffused electron distribution than In$^+$ in an external electric field, indicating that larger basis sets are required for computing Sr. For Sr, the uncertainty in $\alpha^J$ is around 1.2$\sim$3.1, most of which is from the error in $P_{SD}$. This reflects that the accurate calculation of $\alpha$ of Sr replies greatly on the basis set. The values of $\gamma^J$ for the $5s^2$ $^1S_0$ and $5s5p$ $^3P^{\rm o}_{0}$ states of Sr, calculated by using the relativistic CC method, are summarized in Table~\ref{table5}. As a higher-order property, the final data of $\gamma^J$ for the $5s^2$ $^1S_0$ and $5s5p$ $^3P^{\rm o}_{0}$ states of Sr show a large dependence on $\Delta P_T$ and $\Delta P_{\rm core}$, and also show larger uncertainties than that for $\alpha^J$.

The previously reported dipole polarizabilities of Sr are also listed in Table~\ref{table4} for comparison. They include the calculation of Sahoo et al. using the relativistic couple cluster (RCC) method \cite{Sahoo-PRA-2008}, the calculation of Porsev et al. using the CI method with many-body perturbation theory (CI+MBPT)~\cite{Porsev-PRA-2008}, the calculation of Safronova et al. using the CI + all-order method \cite{Safronova-PRA-2013-Sr}, the calculation of Mitroy et al. using the CI method with semiempirical core polarization potential (CICP) \cite{Mitroy-MP-2010}, and the calculation of Porsev et al. using CI + all-order method \cite{Porsev-PRA-2014}. Noteworthy to mention are the CI+all-order results of Safronova et al. that give $\alpha$=194.4 and 441.9 for $5s^2$ $^1S_0$ and $5s5p$ $^3P^{\rm o}_{0}$, respectively. For such two states, we obtain $\alpha^J$=199.7 and 444.1 in the relativistic CC calculation, which are slightly larger than the values of Safronova et al. The possible reason for such discrepancies may be due to underestimation of the magnitudes of $\Delta P_T$ and $\Delta P_{\rm core}$ for which our basis sets are much smaller. There are fewer reported data available for the $5s5p$ $^3P^{\rm o}_{1}$ state. Porsev et al. have obtained $\alpha^{J}$=498.8 for the $5s5p$ $^3P^{\rm o}_{1}$ state with $M_J=1$ state by using the CI+MBPT method, which is close to our calculated result $\alpha^{J}$=480.9. The scalar and tensor polarizabilities of $5s5p$ $^3P^{\rm o}_{1}$ have also been given as $\bar{\alpha}^J$=497.0 and $\alpha_a^J$=27.7 by Mitroy et al. \cite{Mitroy-MP-2010} and $\bar{\alpha}^J$=459.2 and $\alpha_a^J$=26.0 by Porsev et al. \cite{Porsev-PRA-2014}. The experimental value of $\alpha_a^J$ for $5s5p$ $^3P^{\rm o}_{1}$ is 24.5. By using the relativistic CI method we obtain $\bar{\alpha}^J$=409.7 and $\alpha_a^J$=23.0 for Sr $5s5p$ $^3P^{\rm o}_1$, which differs from Porsev's values by 11\%. Such a relativistic CI calculation is implemented with much smaller basis set and has about 10\% error in $\alpha^J$ in comparison with our relativistic CC calculation. Considering this error, our results are much closer to the values of Porsev et al.~\cite{Porsev-PRA-2014}. There is no reported data available for the dipole polarizability of the Sr $5s5p$ $^3P^{\rm o}_2$ state. In the present relativistic CC calculations we obtain $\bar{\alpha}^J$=491.1 and $\alpha_a^J$=--54.5 for $5s5p$ $^3P^{\rm o}_2$. For the hyperpolarizabilities of Sr we recommend $\gamma^J$=691957 and 3228219 for $5s^2$ $^1S_0$ and $5s5p$ $^3P^{\rm o}_{0}$, respectively.

The results of $\alpha^L$ for the states of $5s^2$ $^1S$ and $5s5p$ $^3P^{\rm o}$ of Sr are given in Table~\ref{table6}. In the scalar relativistic CI calculations, $\alpha^L$ for the $5s^2$ $^1S$ ground state seems to be insensitive to the basis set, $\Delta P_Q$, and $\Delta P_{\rm core}$; on the other hand, $\alpha^L$ for the $5s5p$ $^3P^{\rm o}$ state is sensitive to these factors. $\alpha^L$ of the $5s5p$ $^3P^{\rm o}$ state increases with the size of basis set and it reaches convergence at $X$=4$\zeta$. $\Delta P_T$ and $\Delta P_{\rm core}$ are substantial in magnitude but they are in opposite sign, resulting in a cancellation with each other in the final value of $\alpha^L$. Notice that this phenomenon is also found in Table~\ref{table4} for the relativistic CC calculation but the cancelation is not as strong as that in Table~\ref{table6} for the scalar relativistic CI calculation. Combining the relativistic CC and scalar relativistic CI results, we can conclude that the basis set has very important effect on the final values of $\alpha^J$ and $\alpha^L$. A further expansion of the basis set will no double improve the accuracy of $\alpha$. As shown in Tables~\ref{table4} and \ref{table6}, for both the relativistic CC and the scalar relativistic CI calculations, $\Delta P_{\rm core}$ has reached convergence upon inclusion of $3d4s4p$ into the electron correlations. This indicates that the minimum number of the core electrons is 18 for the case of Sr, i.e., the correlation contribution from $3d4s4p$ needs to be included in all-electron calculation in order to achieve an accurate dipole polarizability.

The relativistic effect in the four-component relativistic formalism can be understood as a combination of spin-orbit coupling effect and contraction or decontraction of the radial electron density, i.e., the scalar relativistic effect. Among the $5s5p$ $^3P^{\rm o}$ states, the $^3P^{\rm o}_0$ component has a spherically symmetric electronic density. In this case, the effect of the pure spin-orbit interaction on the polarizability can be reflected by comparing the results obtained by fully and scalar relativistic calculations for the $^3P^{\rm o}_0$ component, as we have done for Al$^+$, in terms of
\begin{equation}
\frac{\bar{Q}^L(^3P^{\rm o})-Q^J(^3P^{\rm o}_0)}{Q^J(^3P^{\rm o}_0)}.
\label{eq:t9}
\end{equation}
Such a fractional difference is about 8\% in $\alpha$ and 22\% in $\gamma$ for the $5s5p$ $^3P^{\rm o}$ states of In$^+$, showing an evident spin-orbit effect in $\alpha$ and $\gamma$. The fractional difference $[\bar{\alpha}^L-\alpha^J(^3P^{\rm o}_0)]/\alpha^J(^3P^{\rm o}_0)$ amounts to 2\% for Sr, which is smaller than that of In$^{+}$. This implies that the effect of the spin-orbit coupling on $\alpha$ of Sr is weaker than In$^{+}$. The relativistic effect can be elucidated by the fractional difference as follows,
\begin{eqnarray}
\frac{\bar{Q}^J(^3P^{\rm o}_1)-Q^J(^3P^{\rm o}_0)}{Q^J(^3P^{\rm o}_0)}
\label{eq:t10}
\end{eqnarray}
and
\begin{eqnarray}
\frac{\bar{Q}^J(^3P^{\rm o}_2)-Q^J(^3P^{\rm o}_0)}{Q^J(^3P^{\rm o}_0)}\,,
\label{eq:t11}
\end{eqnarray}
where $Q$ stands for either $\alpha$ or $\gamma$. Such two fractional differences are about 8.9\% and 9.6\% for $\alpha$ and 28\% and 23\% for $\gamma$ in the case of In$^{+}$. Such variations of $\bar{\alpha}^J$ for different $J$ components for Sr, as reflected by Eqs.~(\ref{eq:t10}) and (\ref{eq:t11}), are about 3.9\% using the values in Refs. \cite{Porsev-PRA-2014} and \cite{Safronova-PRA-2013-Sr}, and 10.5\% using our values. The above values of the fractional differences of $5s5p$ $^3P^{\rm o}_1$ and $^3P^{\rm o}_2$ with respect to $5s5p$ $^3P^{\rm o}_0$ are substantial that reveal the important role of the relativistic effect on $\alpha$ and $\gamma$ for the states of $5s5p$ $^3P^{\rm o}$ of both In$^+$ and Sr.

\subsection{Quadrupole moment and polarizability}
Table~\ref{table7} presents $\theta$ and $\alpha_2$ for the $5s^2$ $^1S$ and $5s5p$ $^3P^{\rm o}$ states of In$^+$ and Sr. The $L$-resolved  $\theta^L$ and $\alpha_2^L$ are obtained by the scalar relativistic calculations (core10)SD(2in4)SDTQ for In$^{+}$ and (core8)SD(2in4)SDTQ for Sr, and the $J$-resolved $\theta^J$ and $\alpha_2^J$ are obtained by the relativistic CC calculations (core10)SD for In$^{+}$ and (core8)SD for Sr. The data for the $5s5p$ $^3P^{\rm o}_1$ state are obtained by the relativistic CI calculations (core10)SD(2in4)SD$<$2 for In$^+$ and (core8)SD(2in4)SD$<$2 for Sr. All calculations are carried out with the $X$=$3\zeta$ basis set. From Table~\ref{table7}, we can found that, the values of $\theta^L$ and $\theta^J$ for the $5s5p$ $^3P^{\rm o}$ state, as well as the values of $\alpha_2^L$ and $\alpha_2^J$ for the $5s^2$ $^1S$ state, are in good accord, which indicates that these properties are insensitive to the spin-orbit coupling. In contrast, the values of $\alpha_2^L$ and $\alpha_2^J$ for the $5s5p$ $^3P^{\rm o}$ state show clear discrepancy, and $\alpha_2^J$ changes greatly for different $J$ components. This means that the spin-orbit coupling plays an important role on the quadrupole polarizabilities of the  $5s5p ^3P^{\rm o}$ states of In$^+$ and Sr.

The quadrupole moments of Sr in the $5s5p$ $^3P^{\rm o}$ state were calculated in earlier works by using the CI+MBPT method \cite{Derevianko-PRL-2003} and the CICP method \cite{Mitroy-PRA-2004}. Our results agree with their data within 1\%. The quadrupole polarizabilities for the $5s^2$ $^1S$ and $5s5p$ $^3P^{\rm o}$ states of Sr were calculated using the CICP method \cite{Mitroy-MP-2010} and the CI+all-order method with the random phase approximation (RPA) applied \cite{Porsev-PRA-2014}. For the ground state $5s^2$ $^1S$, our results are in agreement with their data within 1\%. For the $5s5p$ $^3P^{\rm o}$ states, our $L$-resolved values agree with the CICP data at the level of 2\%, and our $J$-resolved result gives $\theta^J$=1.17$\times10^5$ for the Sr $5s5p$ $^3P^{\rm o}_1$ state, in consistent with the CI+all-order+RPA value 1.05$\times10^5$ \cite{Porsev-PRA-2014} within 11\%.

\subsection{Black-body Radiation Shift}

Finally, our results for the dipole polarizabilities, the second dipole polarizabilities, and the quadrupole polarizabilities are used to estimate the BBR shift in the clock transition frequency of In$^{+}$ and Sr. The BBR shift can be written in the form \cite{Porsev-PRA-2006, Arora-PRA-2012}
\begin{equation}
\delta E_{\rm BBR}=-\frac{1}{2}\Delta \alpha \langle E^2_{E1}\rangle-\frac{1}{24}\Delta \gamma \langle E^2_{E1}\rangle ^{2}-\frac{1}{2}\Delta \alpha_{2} \langle E^2_{E2}\rangle,
\label{eq:t12}
\end{equation}
where $\langle E^2_{E1} \rangle$ and $\langle E^2_{E2} \rangle$ are the averaged electric fields induced by the electric dipole E1 and the electric quadrupole E2 and they are respectively
\begin{eqnarray}
\langle E^2_{E1}\rangle=\frac{4\pi^3\alpha_{\rm fs}^3}{15}(\frac{k_BT}{E_h})^4
\label{eq:t13}
\end{eqnarray}
and
\begin{eqnarray}
 \langle E^2_{E2}\rangle=\frac{8\pi^5\alpha_{\rm fs}^5}{189}(\frac{k_BT}{E_h})^6\,.
\label{eq:t14}
\end{eqnarray}
In the above, $\alpha_{\rm fs}$ is the fine structure constant, $k_BT/E_h\approx 10^{-9}$ for the temperature $T$=300~K, $k_B$ is the Boltzmann constant, $E_h$ is the Hartree energy, and $\Delta \alpha$, $\Delta \alpha_{2}$ and $\Delta \gamma$, expressed in atomic units, are respectively the differences of the dipole polarizability, quadrupole polarizability and dipole hyperpolarizability between $5s^2$  $^1S_0$ and $5s5p$ $^3P^{\rm o}_0$ of In$^+$ and Sr.
In Eq.~(\ref{eq:t12}) we have neglected the dynamic fractional correction to the total shift \cite{Porsev-PRA-2006} and assume that the contribution of the hyperpolarizability to the BBR shift can be approximated by the AC-stark shift $\langle E^2_{E1}\rangle^2$ for a given electric field. Using the data of $\alpha$, $\gamma$, and $\alpha_2$ obtained by the relativistic CC calculations in Tables~\ref{table1}, \ref{table2}, and \ref{table7} for the states of $5s^2$ $^1S_0$ and $5s5p$ $^3P^{\rm o}_0$ of In$^+$ , the BBR shifts due to $\alpha$, $\alpha_{2}$, and $\gamma$ are determined to be 0.017, 8.33$\times10^{-10}$, and 1.93$\times10^{-17}$ Hz, respectively, for the In$^{+}$ clock transition frequency; and using the data of $\alpha$, $\gamma$, and $\alpha_2$ obtained by the relativistic CC calculations in Tables~\ref{table4}, \ref{table5}, and \ref{table7} for the states of $5s^2$  $^1S_0$ and $5s5p$ $^3P^{\rm o}_0$ of Sr, the BBR shifts due to $\alpha$, $\gamma$, and $\alpha_{2}$ are determined to be 2.09 and 5.82$\times10^{-8}$,  and 1.69$\times10^{-15}$ Hz, respectively, for the Sr clock transition frequency. It is clear that the contributions from  $\gamma$ and $\alpha_2$ are far less important than that from $\alpha$ in the BBR shifts and can thus be safely omitted according to the current precision of the quoted 10$^{-18}$ uncertainty of the time frequency standard.

\section{Conclusion}
In summary, we have calculated $\alpha$, $\gamma$, $\theta$, and $\alpha_2$ for the ground state $5s^2$ $^1S_0$ and the low-lying excited states $5s5p$ $^3P^{\rm o}_{0,1,2}$ of In$^+$ and Sr by using the finite field method. A satisfactory accuracy is achieved through convergence studies for the basis sets and sufficient inclusion of the electron correlations. This method can also be applied with similar accuracy to the calculations of the polarizabilities of atomic cores, see Supplemental Material at [URL will be inserted by publisher] for $\alpha$, $\gamma$, and $\alpha_2$ of In$^{3+}$ and Sr$^{2+}$. Thus, it will be useful to employ the finite field method to perform a fast evaluation of required properties, especially when high-precision experimental studies and sophistical sum-over-state calculations are not available or all available results are not in complete agreement. It is noteworthy to mention that
the errors of the finite field calculations need to be examined carefully and minimized for such applications, which requires a detailed knowledge about the rate of convergence of basis set and electron correlations for a property of interest.

In our finite field calculations, we have investigated the influences of the basis set and the level of electron correlations on the computed properties. For the case of In$^{+}$, the convergence for $\alpha$ can be easily reached when the basis set is increased up to $X$=$4\zeta$ and the core electrons included into the correlation calculation are increased up to $3d4s4p4d$. The dominant correction to $\alpha$ is from $\Delta P_T$, and thus a more accurate evaluation of the contribution of the triple excitation is needed for higher accuracy. For example, the relativistic CCSDT calculation with the 4$\zeta$ basis set may give more accurate results, although such a study is prohibited at this moment given the present computing resources.
For the case of Sr, the convergence of $\alpha$ is highly sensitive to the quality of the basis set because a neutral atom has more diffused electron density in an external field than a positive charged ion. In this case, the predominant factor will be the expansion and optimization of the basis sets beyond $X$=4$\zeta$. The higher-order corrections of the quadruple excitation of the electron correlation and the Breit and QED corrections should be roughly two- and one-order of magnitude smaller than $\Delta P_T$ in the relativistic CC calculations and accordingly the resulted uncertainties should be at the level of 0.01\% and of 0.1\%, which are very small.

Our fully relativistic calculations have shown that the $5s5p$ $^3P^{\rm o}_{0,1,2}$ states of In$^{+}$ and Sr have obviously different values of polarizabilities, which reflects the important contributions of the relativistic and spin-orbit coupling interactions. Some general trends about the sole effect of the spin-orbit coupling are worth noticing through comparative studies using the fully and scalar relativistic approaches. The fractional difference between the $L$- and $J$-resolved $\alpha$ of $5s5p$ $^3P^{\rm o}_{0}$ is about 8\% for In$^{+}$, but only 2\% for Sr, implying that the effect of the spin-orbit interaction on $\alpha$ of In$^{+}$ is stronger than Sr.

\section{Acknowledgements}
The authors would like to thank Prof. Zong-Chao Yan for careful reading and revising the manuscript and to Dr. Jun Jiang and Dr. Chengbin Li for valuable comments. This work is supported by 2012CB821305, NSFC 61275129, NFSC 21203147, and CAS KJZD-EW-W02.

\section{Author contribution statement}
All authors contributed equally to the paper.

\newpage
\end{document}